\documentclass[twocolumn,aip,apl,floatfix,reprint]{revtex4}

\usepackage{graphicx}
\usepackage{epstopdf}
\usepackage{amsmath}
\usepackage{latexsym}
\usepackage{amsmath}
\usepackage{esdiff}
\usepackage[final]{pdfpages}

\newcommand{\beq}{\begin{equation}}
\newcommand{\eeq}{\end{equation}}

\begin{document}

\title{First principles study of hBN-AlN short-period superlattice heterostructures}

\author{Catalin D. Spataru}
\thanks{Corresponding author, cdspata@sandia.gov.}
\affiliation{Sandia National Laboratories, Livermore, California 94551, USA}
\author{Mary H. Crawford}
\author{Andrew A. Allerman}
\affiliation{Sandia National Laboratories, Albuquerque, New Mexico 87123, USA}

\begin{abstract}
We report a theoretical study of the structural, electronic and optical properties of hBN-AlN superlattice heterostructures (SL) using a first-principles approach based on standard and hybrid Density Functional Theory. We consider short-period ($L<10$ nm) SL and find that their properties depend strongly on the AlN layer thickness $L_{AlN}$.
For $L_{AlN}\lesssim1$ nm, AlN stabilizes into the hexagonal phase and SL display insulating behavior with type II interface band alignment and optical gaps as small as $5.2$ eV. The wurtzite phase forms for thicker AlN layers.
In these cases built-in electric fields lead to formation of polarization compensating charges as well as two-dimensional conductive behavior for electronic transport along interfaces. We also find defect-like states localized at interfaces which are optically active in the visible range.
\end{abstract}

\maketitle

Group III-nitride ultraviolet light emitting diode (UV-LED) technologies possess unique properties that enable key applications in sensing,  communication and optoelectronics \cite{UVemitters_book}. Significant efforts worldwide have developed AlGaN-based alloys for UV-LEDs, however fundamental materials challenges have precluded high-performing devices to date. Remaining issues include insufficient p-type doping in optically transparent layers due to fundamentally high acceptor activation energies \cite{Shatalov}.

Hexagonal boron nitride (hBN), known as the ideal subtrate for high-mobility graphene, has recently emerged as a highly promising candidate UV material \cite{Chubarov}, possessing a large band gap of $\sim6$ eV \cite{Cassabois,MajetyJiang} yet remarkably showing effective p-type doping \cite{Majety_hBN,hBN_p_doping}. Exploring hB(Al,Ga)N alloys represents a possible avenue for realizing efficient, tailorable UV-LEDs and photodetectors. However, given the difference between the 
tetrahedral coordination/$sp^3$ hybridization in wurtzite AlN or GaN and the trigonal planar coordination/$sp^2$ hybridization in hBN, one expects high defect formation energy for Al,Ga substituting B and alloy formation may be difficult to achieve.  

hBN-AlN and hBN-GaN superlattice heterostructures offer an alternative route towards tailorable optical and transport properties. Here we present a theoretical study that explores the potential of hBN-AlN superlattices (SL) by predicting their structural, electronic and optical properties via a first-principles approach. Several important questions can be raised. It is well known that built-in electric fields lead to tunable band gaps in GaN/AlN superlattices \cite{StampflPRB10}. Do these effect exist in SL? Also, first principles studies show that unsupported AlN layers may undergo a wurtzite to hexagonal phase transformation  \cite{FreemanPRL06,Baca15}. Do these transformations occur in SL? If yes, what are the implications for the SL electronic and optical properties?
We answer these questions via {\it ab initio} Density Functional Theory (DFT) \cite{DFT} calculations 
using standard and hybrid DFT functionals. The calculations are performed with the code VASP \cite{Kresse} in conjunction with projector augmented wave pseudo-potentials \cite{PAW}.

\begin{figure}
\vspace{-0.0cm}
\begin{center}
{\includegraphics[trim=10 10 40 20,clip,width=\columnwidth]{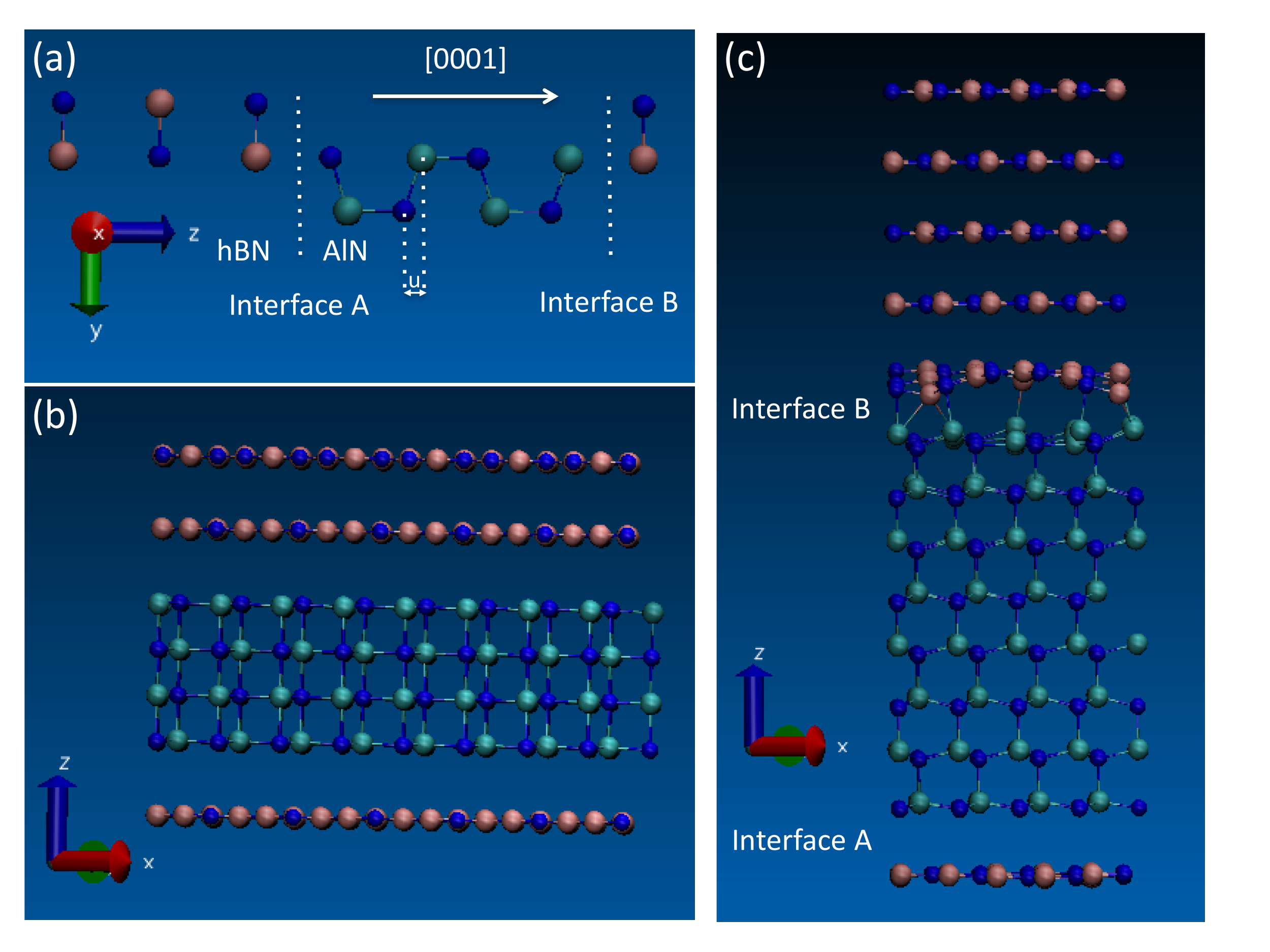}}
\end{center}
\caption{a) Illustration of the starting configuration (before optimization) of the $(hBN)_3(AlN)_4$ superlattice. Side view of the optimized supercell geometry of: b) $6\sqrt{3}\times6\sqrt{3}/8\times8$  $(hBN)_3(AlN)_4$ SL, c) $5\times5/4\times4$  $(hBN)_6(AlN)_8$ SL.}
\vspace{-0.5cm}
\label{sketch_SL}
\end{figure}

We obtain SL structures by aligning nm-scale layers of hBN (AA' stacking) and AlN along the [0001] direction (the growth direction for the bulk materials). This is illustrated in Fig. \ref{sketch_SL} for the particular case of $(hBN)_3(AlN)_4$ where the subscripts denote the number of hBN monolayers and AlN bilayers in each of the two layers. The initial configurations (before optimization) accommodate AlN in the wurtzite phase (wAlN). There are two inequivalent interfaces between the two layers depending whether the surface of the AlN layer terminates with N (interface A) or Al (interface B).

The lattice constants $a$ of hBN and wAlN show a significant mismatch along directions parallel to the SL interface (see Table \ref{table1}), posing a challenge to realistic {\it ab initio} simulations which require large supercells.  
We construct the supercell by replicating the surface unit cells for each layer until the hBN/AlN supercells can be matched with less than $\sim1\%$ tensile/compressive lateral strain \cite{Stokbro,GohdaAPL12}. 
After atomic optimization we find that ultrashort AlN layers ($L_{AlN}\lesssim 1$ nm) relax into the hexagonal phase, while thicker layers remain in the wurtzite phase.
For AlN in wurtzite phase the matching condition is satisfied with $5\times5$ hBN and $4\times4$ AlN surface unit cells sharing the same rotational orientation. For AlN in the hexagonal phase a larger supercell is needed with hBN rotated $30^{\circ}$ w.r.t AlN and $6\times6$ hBN matching $8\times8$ AlN lateral replicas.

\begin{table}
\vspace{-0.25cm}
\caption{\label{table1} {\it Ab initio} results of AlN (wurzite and hexagonal phase) and hBN bulk properties. Optimized lattice constants $a$ and $c$ of the hexagonal primitive unit cell are obtained within LDA, band gaps are obtained within hybrid-DFT (HSE06).}
\begin{ruledtabular}
\begin{tabular}{ccccc}
& a [\AA] & c [\AA] & Minimum gap [eV] & Direct gap [eV]\\
\hline
wAlN & 3.08 & 4.92 & 5.91 & 5.91\\
hAlN & 3.28 & 4.10 & 4.91 & 4.91\\
hBN & 2.49 & 6.50 & 5.56 & 6.05\\
\end{tabular}
\end{ruledtabular}
\vspace{-0.5cm}
\end{table}

We use a standard DFT functional in the  local density approximation (LDA) \cite{LDA} to optimize the atomic structure of the SL, relaxing the atomic forces to better than $0.01$ eV/\AA\ accuracy. In general standard DFT functionals greatly underestimate the band gaps of materials \cite{bandgapsDFT}, in particular that of  superlattices \cite{StampflPRB10}. In order to obtain realistic SL electronic and optical properties we include exchange-correlation effects via the hybrid-DFT HSE06 \cite{HSE06} functional ($0.25$ mixing parameter). We use an energy cutoff of $500$ eV. Convergence w.r.t. k-point sampling of the Brillouin zone is achieved with Monkhorst-Pack \cite{Monk} grids ranging from 2x2x2 (shorter SL) to 4x4x1 (larger SL). 

We analyze the structure of the AlN layer after optimization by considering the distance $u$ along the z-axis between Al and N atomic sheets within a given AlN bilayer (see fig. \ref{sketch_SL}(a)). The value of $u$ for each of the SL ($u_{SL}$), normalized to the value of u for bulk wAlN ($u_{wAlN}=0.58$ \AA), defines an internal structure parameter $s\equiv u_{SL}/u_{wAlN}$ that equals $0$($1$) for the pure hexagonal(wurtzite) phase. After optimization and near the middle of the AlN layer we find $s=0$ for $(hBN)_6(AlN)_4$, $(hBN)_3(AlN)_4$, $(hBN)_3(AlN)_2$ and $(hBN)_1(AlN)_4$, $s=0.78$ for $(hBN)_6(AlN)_8$ and $s=0.91$ for $(hBN)_{12}(AlN)_{16}$ and $(hBN)_6(AlN)_{16}$ SL. This suggests that the structure of the AlN layer in the optimized SL depends mainly on the AlN layer thickness $L_{AlN}$, with ultrashort/thicker AlN layers relaxing into the hexagonal/wurtzite phase. We focus next on symmetric SL having a ratio between hBN and AlN layer thickness $L_{hBN}/L_{AlN} \approx 1$ and a total length period $L=L_{hBN}+L_{AlN}$ ranging from about $2$ to $8$ nm.

First, we consider the ultrathin, unrotated $(hBN)_3(AlN)_4$ $5\times5/4\times4$ configuration with $278$ atoms/supercell and $L_{AlN}\sim1$ nm. Upon optimization AlN relaxes into the hexagonal phase (hAlN) even though this transition must overcome $\sim5\%$ mismatch strain (see table \ref{table1}). We also consider the $30^{\circ}$ rotated configuration $6\sqrt{3}\times6\sqrt{3}/8\times8$, containing 1060 atoms/supercell. The optimized structure is seen in Fig. \ref{sketch_SL}(b) showing the AlN layer in the hexagonal phase with an interlayer distance of $\approx2.1$ \AA. The AlN slab is non-polar and the two interfaces are equivalent subject to only $\sim1\%$ mismatch strain. Both unrotated and rotated configurations lead to the same qualitative picture of the structural, electronic and optical properties of the $(hBN)_3(AlN)_4$ SL. 

Next we consider larger period SL, namely $(hBN)_6(AlN)_8$ and $(hBN)_{12}(AlN)_{16}$. We simulate them using unrotated $5\times5/4\times4$ configurations with $556$ and $1112$ atoms/supercell respectively.
Fig. \ref{sketch_SL}(c) shows the side view of the optimized $(hBN)_6(AlN)_8$ SL with $L_{AlN}\sim2$ nm. The bulk of the AlN layer clearly 
belongs to the wurtzite phase. However, there is significant AlN surface relaxation at interface A (here $s\approx 0.5$) and noticeable structural distortion (random out-of-plane atomic displacements as large as $1$ \AA ) forming across interface B (see fig. \ref{sketch_SL}(c)). 
A similar picture holds for the larger period ($L_{AlN}\sim4$ nm) $(hBN)_{12}(AlN)_{16}$ SL. Corroborated with our findings for the ultrathin SL as well as the other aforementioned non-symmetric SL, this indicates that the critical AlN thickness beyond which the AlN phase changes from hexagonal to wurtzite lies between $1$ and $2$ nm.  

It is not surprising that AlN stabilizes in the hexagonal phase in ultrathin hBN/AlN SL. Indeed, unsupported, ultrashort hAlN layers have been predicted to be more stable than their wurtzite counterpart \cite{FreemanPRL06,Baca15} and evidence of epitaxial growth of thin AlN films
on Ag(111) substrates has recently been reported \cite{Tsipas13}.  
A stabilization mechanism involving global transformation from the wurtzite to hexagonal phase has also been predicted and measured in ultrathin films of several metal oxides  
\cite{NoguerraPRL04,TuschePRL07}. Such transformation results from competing surface and bulk energy terms with the former winning at small thickness \cite{Ch7Noguerra}. 

The adhesion properties of the two layers in the SL can be estimated from the binding energy $E_b$ of the SL defined as: $E_b=(E_{SL}-E_{hBN}^{layer}-E_{AlN}^{layer})/A$,
where $A$ is the interface area, $E_{SL}$ is the total energy of the SL, and $E_{hBN/AlN}^{layer}$ are the total energies of the isolated hBN/AlN layers (obtained via separate slab calculations that include further structural optimization). We find: $E_b=-18$, $-33$ and $-21$ meV/\AA$^2$ respectively for the $(hBN)_3(AlN)_4$, $(hBN)_6(AlN)_8$ and $(hBN)_{12}(AlN)_{16}$ SL. By contrast, the typical average formation energy $E_f$ of the hBN/AlN interface (relative to bulk hBN and AlN \cite{Bernardini}) is higher, dominated by the large AlN surface formation energy. For example, for the $(hBN)_3(AlN)_4$ SL we find that $E_f=62$ meV/\AA$^2$, very close to the calculated formation energy ($65$ meV/\AA$^2$) of the hAlN (0001) surface.

\begin{figure}
\vspace{-0.0cm}
\begin{center}
\resizebox{7.0cm}{!}{\includegraphics[trim=0 0 0 0,clip,width=\columnwidth]{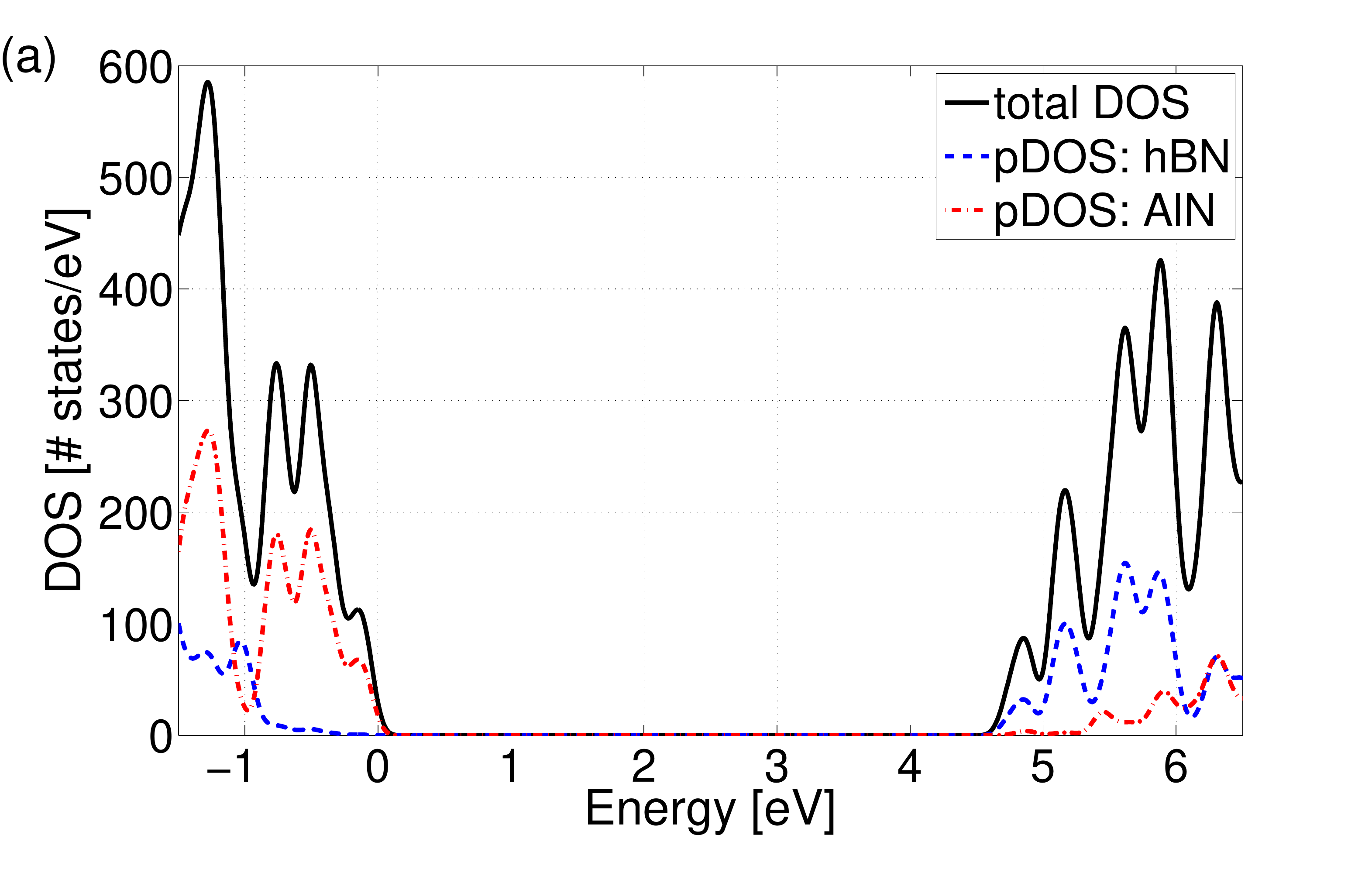}}
\resizebox{7.0cm}{!}{\includegraphics[trim=0 20 0 0,clip,width=\columnwidth]{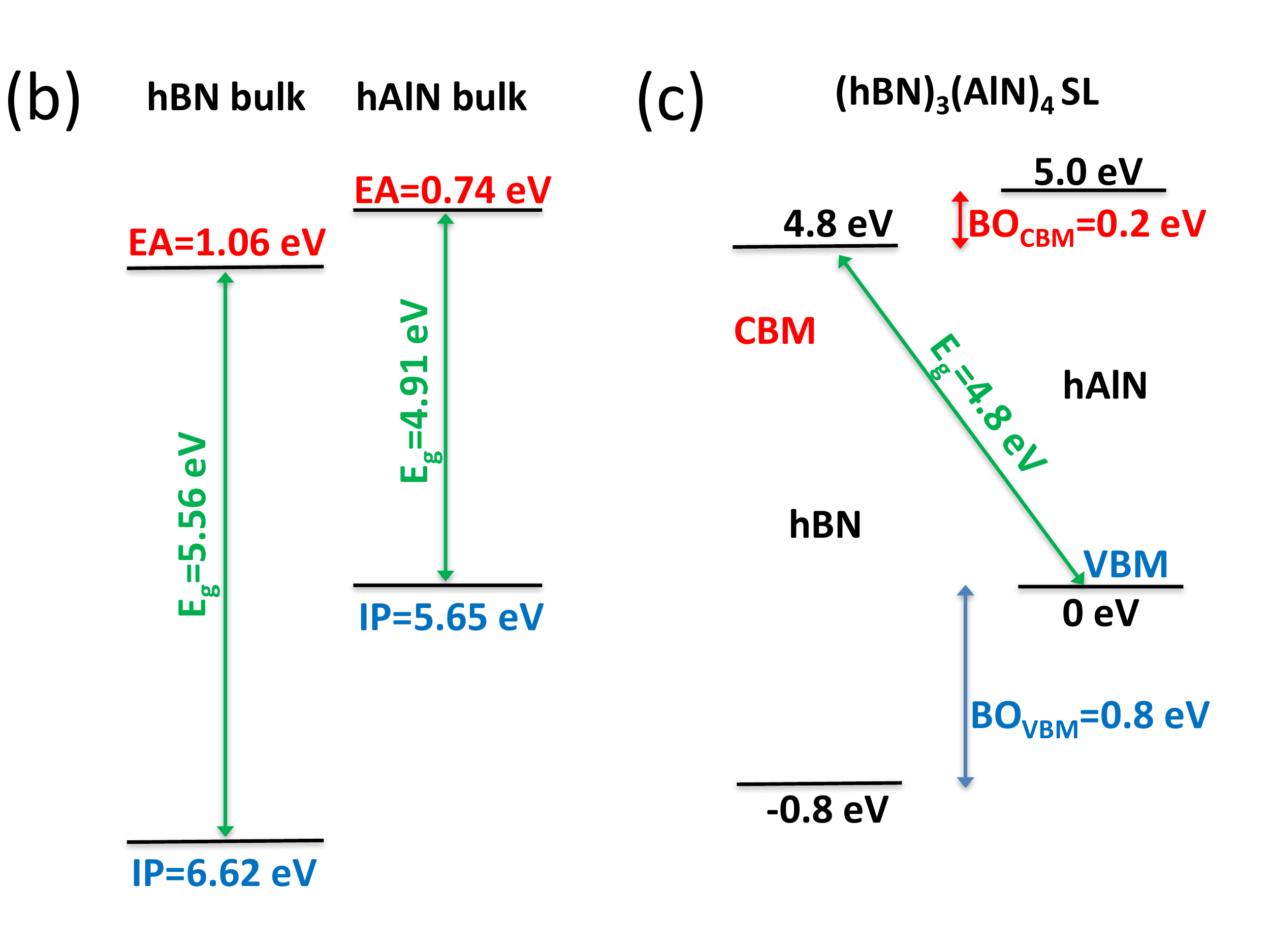}}
\end{center}
\vspace{-0.5cm}
\caption{a) DOS: total and projected on individual layers (hBN or AlN) of the $(hBN)_3(AlN)_4$ SL. Zero energy set by VBM. b) Illustration of natural band alignment for bulk hBN and bulk hAlN. EA and IP referenced w.r.t. the vacuum level. c) Illustration of band offsets (BO) and band gaps ($E_g$) for the $(hBN)_3(AlN)_4$ SL. Zero energy set by VBM. Results obtained within hybrid-DFT (HSE06).}
\vspace{-0.75cm}
\label{dos_ultra}
\end{figure}

As a consequence of the AlN global phase transformation the electronic properties of SL depend dramatically on $L_{AlN}$. Fig. \ref{dos_ultra} shows the calculated density of states (DOS) of the $(hBN)_3(AlN)_4$ SL (rotated configuration). The SL is an insulator with a minimum band gap of $4.8$ eV, slightly smaller than that of the individual layers in bulk form (see Table \ref{table1}). The calculated bandstructure (see Fig. S1 in Suppl. Mat.) indicates an indirect bandgap.

The band alignment at the interface of heterostructures is of great interest in opto-electronic applications \cite{Walle87,Baldereschi}. 
Fig. \ref{dos_ultra}b) illustrates schematically the natural band alignment of bulk hBN and bulk hAlN showing the electron affinity (EA) and ionization potential (IP) obtained from hybrid-DFT (HSE06) VBM and CBM eigenvalues referenced to the vacuum level (the latter obtained for each material by aligning the average electrostatic potential between bulk and the middle of an optimized slab - we considered slabs comprised of $6$ hBN and $12$ hAlN atomic sheets). The corresponding band offsets are $\approx 1.0$ eV at VBM and $\approx 0.3$ eV at CBM.
We note the large valence band offset despite the fact that the states near VBM have N $p_z$ character in both hBN and hAlN. We believe this is due to the very large difference in the lattice constants of these systems (as large as $60\%$, see Table \ref{table1}). 

\begin{figure}
\vspace{-0.0cm}
\begin{center}
\resizebox{7.0cm}{!}{\includegraphics[trim=0 50 0 0,clip,width=\columnwidth]{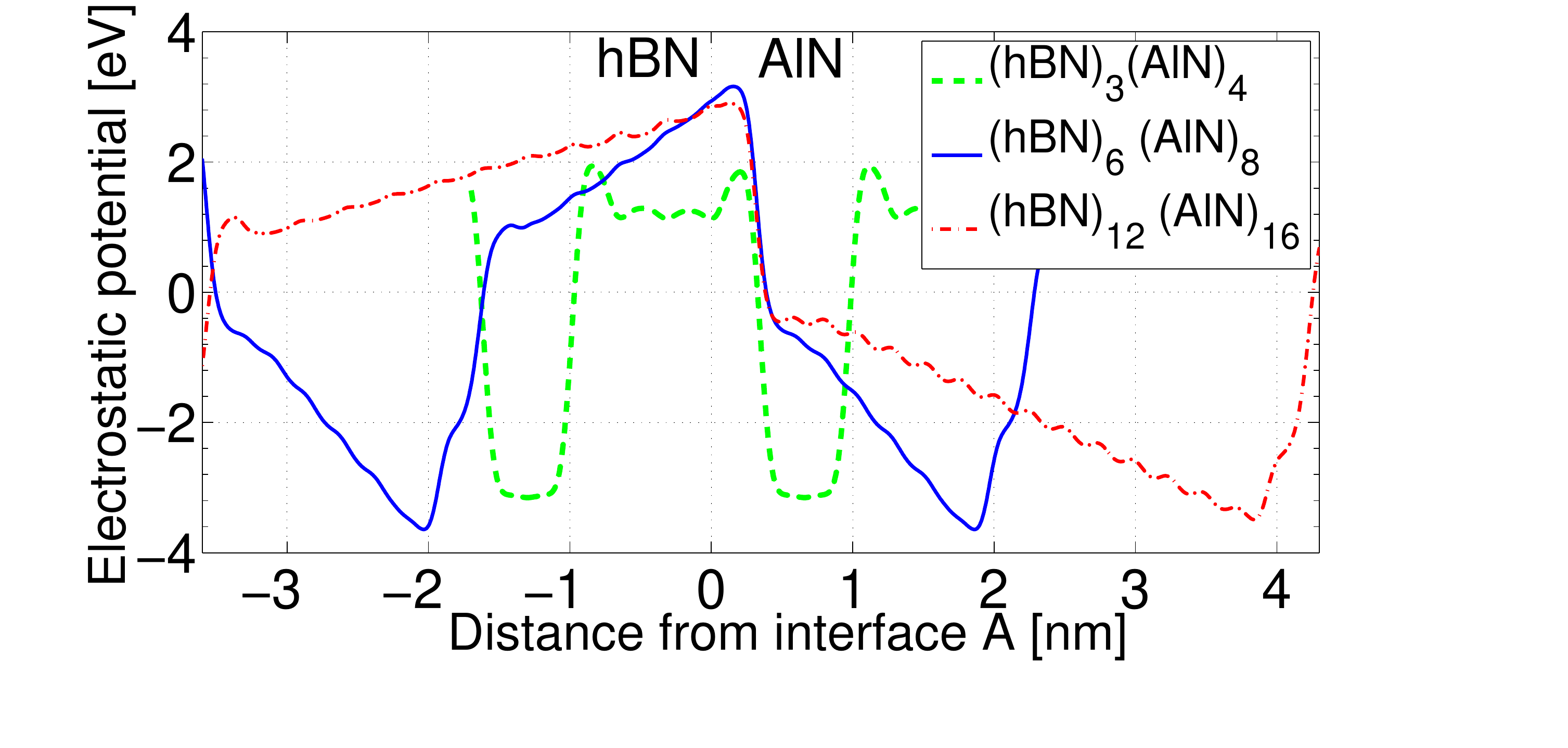}}
\end{center}
\vspace{-0.5cm}
\caption{Macroscopic electrostatic potential profile along [0001] for the $(hBN)_3(AlN)_4$, $(hBN)_6(AlN)_8$ and $(hBN)_{12}(AlN)_{16}$ SL obtained within hybrid-DFT (HSE06).}
\vspace{-0.75cm}
\label{macro_avg}
\end{figure}

By projecting the DOS on individual layers (hBN or AlN) one can deduce the band alignment type in the SL. The projected DOS (pDOS) shown in fig. \ref{dos_ultra}a) indicates that in the $(hBN)_3(AlN)_4$ SL the states near the valence band maximum (VBM) are localized in AlN while those near the conduction band minimum (CBM) localize in hBN. This is characteristic of a type II alignment with staggered gaps, in contrast with type I alignment suggested by a recent  
study \cite{Prete} of monolayer hBN and hAlN based on their bare (no coupling) CBM and VBM energy levels.
Additional analysis based on the layer-projected bandstructure of the $(hBN)_3(AlN)_4$ SL (see figs. S2 and S3 in Suppl. Mat.) shows that the band offsets are $\approx 0.8$ eV at VBM and $\approx 0.2$ eV at CBM, as illustrated in fig. \ref{dos_ultra}c). The difference between these band offsets and those implied by the natural band alignment (fig. \ref{dos_ultra}b)) is due to a combination of charge-transfer interface dipoles, strain and quantum confinement effects. Examination of the planar-averaged electrostatic potential difference between the SL and the sum of the isolated layers indicates that the impact of charge-transfer dipoles is to increase the band offsets by $\approx 0.2$ eV (see fig. S4 in Suppl. Mat.). We have also used the approach described in ref. \cite{Bernardini} to obtain band offsets using hybrid-DFT (HSE06) eigenvalues for bulk hBN and bulk hAlN (referred to the average bulk electrostatic potential) as well as the drop in the macroscopic electrostatic potential across the interface ($\approx 4.4$ eV based on fig. \ref{macro_avg}) of the $(hBN)_3(AlN)_4$ SL. The resulting band offsets are $\approx 1.15$ eV at VBM and $\approx 0.5$ eV at CBM. We ascribe the difference between these values and the band offsets shown in fig. \ref{dos_ultra}c) to strain and finite size effects. 
For comparison we mention the minimum band gaps we calculate using hybrid-DFT (HSE06) for the isolated (fully relaxed within LDA) $(hBN)_3$ and $(hAlN)_4$ slabs, namely $5.68$ and $4.75$ eV respectively.

In the case of $(hBN)_6(AlN)_8$ and $(hBN)_{12}(AlN)_{16}$ SL the difference in polarization (spontaneous plus piezoelectric contributions) between the wAlN slab (polar) and hBN (non-polar) leads to electric fields inside the layers \cite{Ambacher}. Fig. \ref{macro_avg} shows the macroscopic electrostatic potential profile that we obtain via a double-average integral (the two length scales are equal to half the $c$ lattice constant of hBN and wAlN -see Table \ref{table1}) along [0001] of the planar-averaged electrostatic potential \cite{Dandrea&Duke,StrakJAP,StrakPSSC}. 
In the case of the $(hBN)_{12}(AlN)_{16}$ SL the magnitude of the macroscopic electric field is $1.0$ ($0.6$) V/nm inside the AlN (hBN) layer; these values are $\approx2.4$ larger in the case of the  $(hBN)_{6}(AlN)_{8}$ SL.
The maximum/minimum of the potential profile corresponds to interface A/B.
We note sharp changes in the electrostatic potential across each interface. These potential drops are mainly generated by the intrinsic difference between the average crystal potential in the two different compounds \cite{Tung}, but they also have important contribution from other effects including charge transfer interface dipoles \cite{Bernardini} (see also fig. S4 in Suppl. Mat.) or interface-induced structural changes. We also note that the overall potential drop along the layers is quite insensitive to the SL period. This is characteristics of the large thickness regime \cite{Goniakowski} where polarity compensating monopole charges \cite{Bernardini} develop in order to prevent an electrostatic catastrophe. Indeed, for a long enough layer, the potential drop along the layer leads to Fermi level ($E_F$) pinning of surface valence and conduction bands which results in compensating surface charges that prevent further potential drop \cite{Noguera08}. 

\begin{figure}
\vspace{-0.0 cm}
\begin{center}
\resizebox{7.0cm}{!}{\includegraphics[trim=0 0 0 0,clip,width=\columnwidth]{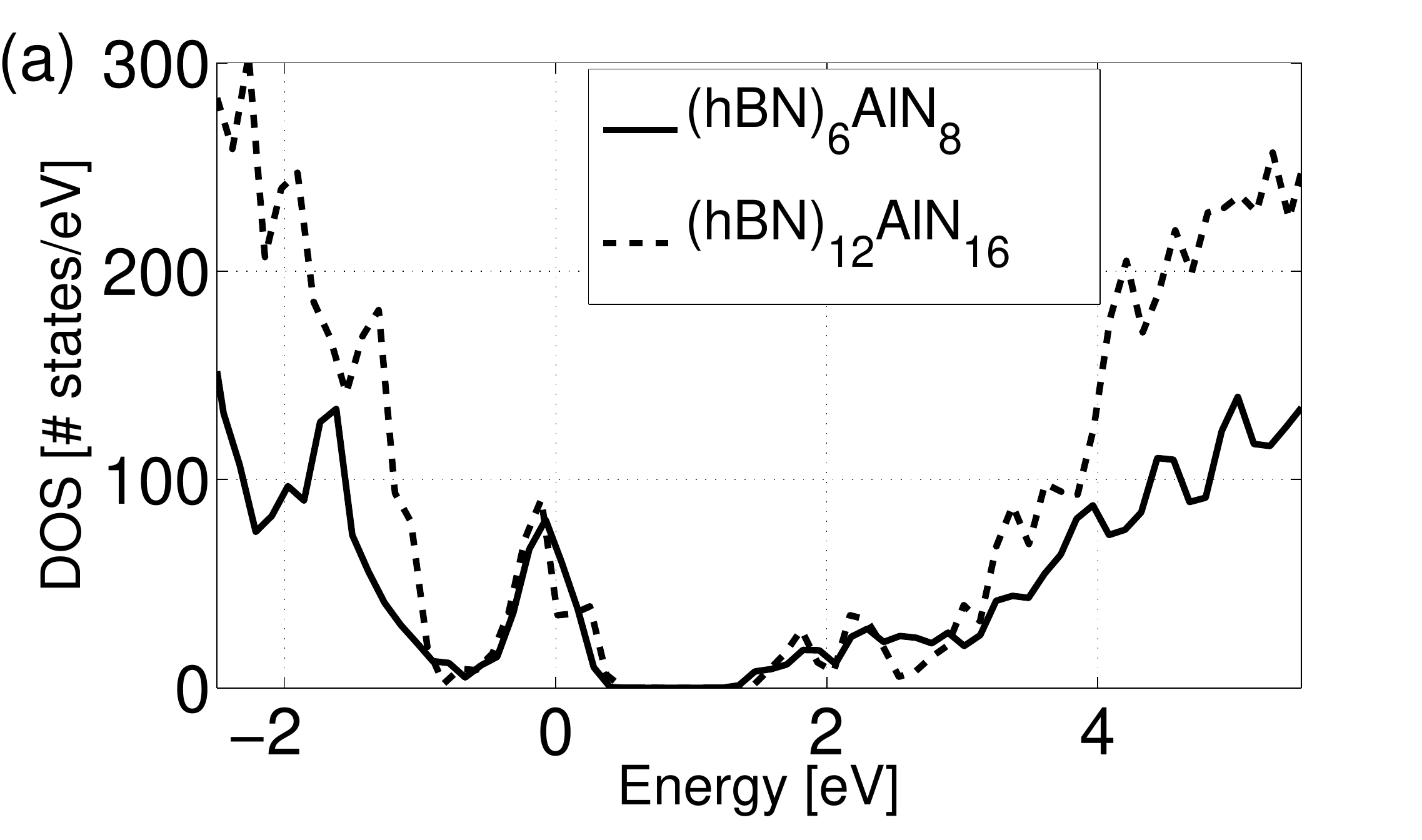}}
\resizebox{7.0cm}{!}{\includegraphics[trim=0 0 10 0,clip,width=\columnwidth]{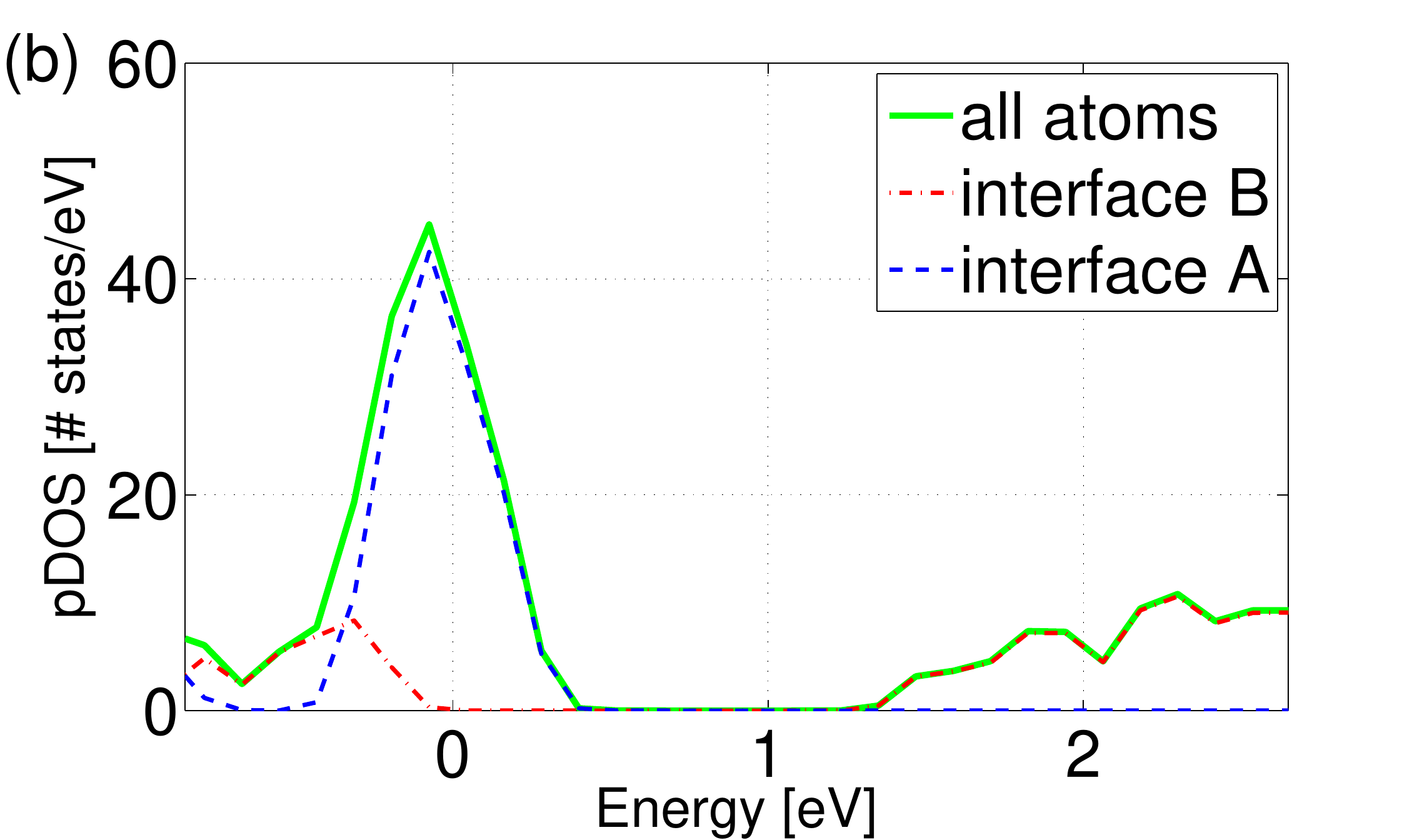}}
\end{center}
\vspace{-0.5 cm}
\caption{(a) Calculated DOS for $(hBN)_6(AlN)_8$ and $(hBN)_{12}(AlN)_{16}$ SL. (b) Orbital projected DOS (pDOS) for $(hBN)_6(AlN)_8$. Green line: all atomic orbitals are included, red (blue) line: only atomic orbitals located in a slab of thickness $1$ nm centered about interface B (A) are included. Results obtained within hybrid-DFT (HSE06).}
\vspace{-0.5 cm}
\label{dos_large}
\end{figure}

Fig. \ref{dos_large}(a) shows the calculated DOS for $(hBN)_6(AlN)_8$ and $(hBN)_{12}(AlN)_{16}$ SL. One notices a prominent feature at $E_F$ which is indicative of electrical conductive behavior for these dopant-free systems.
The corresponding occupied electronic states account for the polarity compensating charges that form in response to the electric field build up. 
Bulk like features start developing at energies  farther away from $E_F$ as indicated by the fact that their weight increases with $L$ -as opposed to features near $E_F$ which vary little with supercell size. Additional information on the atomic character of the DOS for the $(hBN)_{12}(AlN)_{16}$ SL is shown in fig. S5 in Suppl. Mat..

The localization of electronic states about interfaces can be deduced from the DOS projected on atomic orbitals, as shown in fig.\ref{dos_large}(b) for the $(hBN)_6(AlN)_8$ SL. We choose orbitals situated either in the entire supercell (green line) or inside $1$ nm-thick slabs centered about interface A (blue line) or B (red line).  The peak centered about $E_F$  is due to two-dimensional (2D) free charge carriers forming at interface A. 
Additional analysis (see Fig. S6 in Suppl. Mat.) reveals that these states form due to surface valence bands pinned by $E_F$ and that they localize mostly on the AlN side of the interface. 
The electronic states associated with the lower energy pDOS peak (about $0.5$ eV below $E_F$) are accommodated at interface B. These states form when the conduction bands cross $E_F$ (see Fig. S6 in Suppl. Material) due to the polarization-induced electric field. In this case, atomic disorder present on both sides of interface B prevents the states from delocalizing laterally. Furthermore, exchange-correlation effects (captured within HSE06) strongly affect these defect-like states pushing the unoccupied states away from the occupied ones, leading to a significant gap separation of $\sim1.5$ eV.

\begin{figure}
\vspace{-0.0 cm}
\begin{center}
\resizebox{7.0cm}{!}{\includegraphics[trim=0 100 0 0,clip,width=\columnwidth]{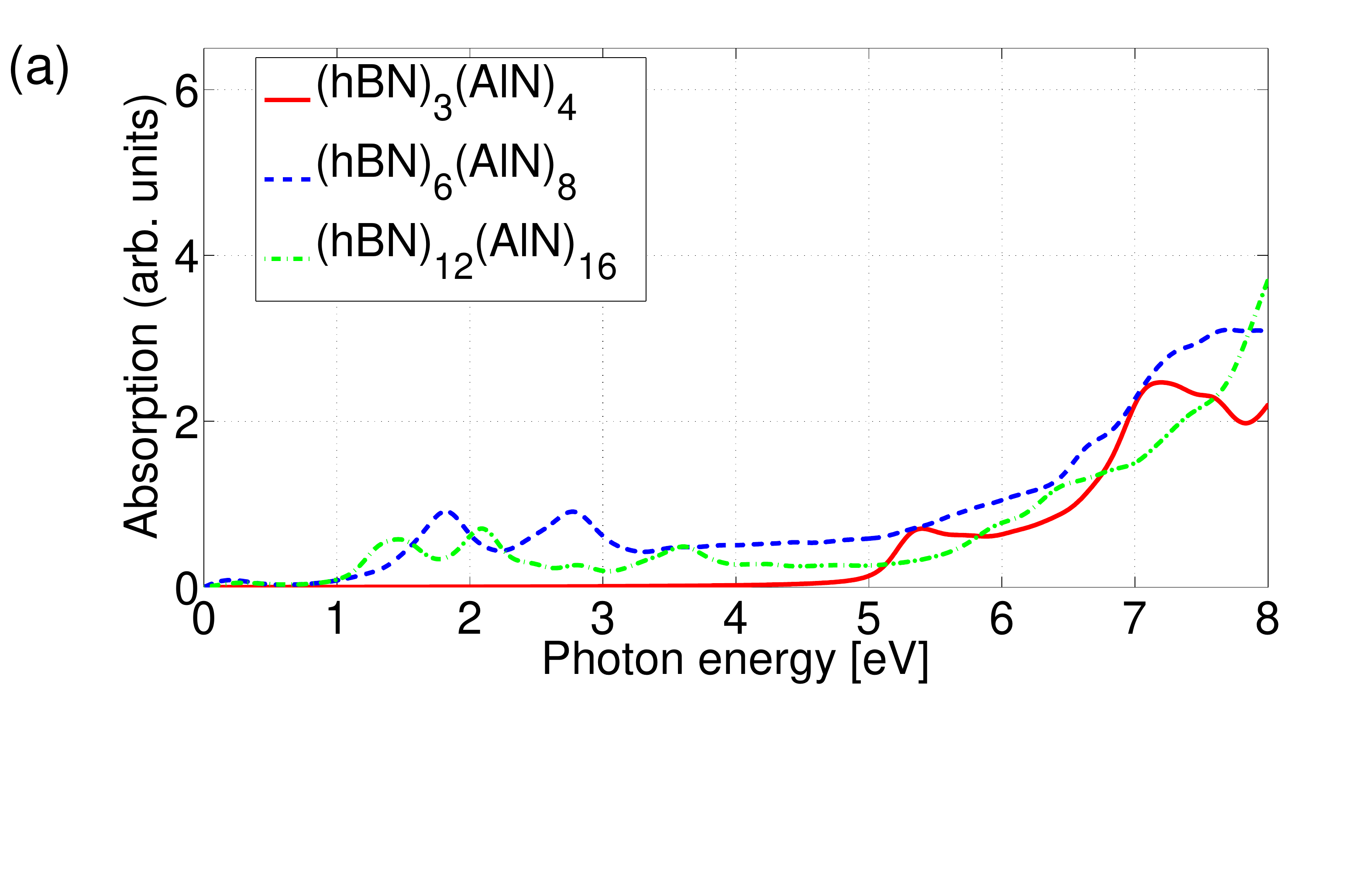}}
\resizebox{7.0cm}{!}{\includegraphics[trim=0 100 0 0,clip,width=\columnwidth]{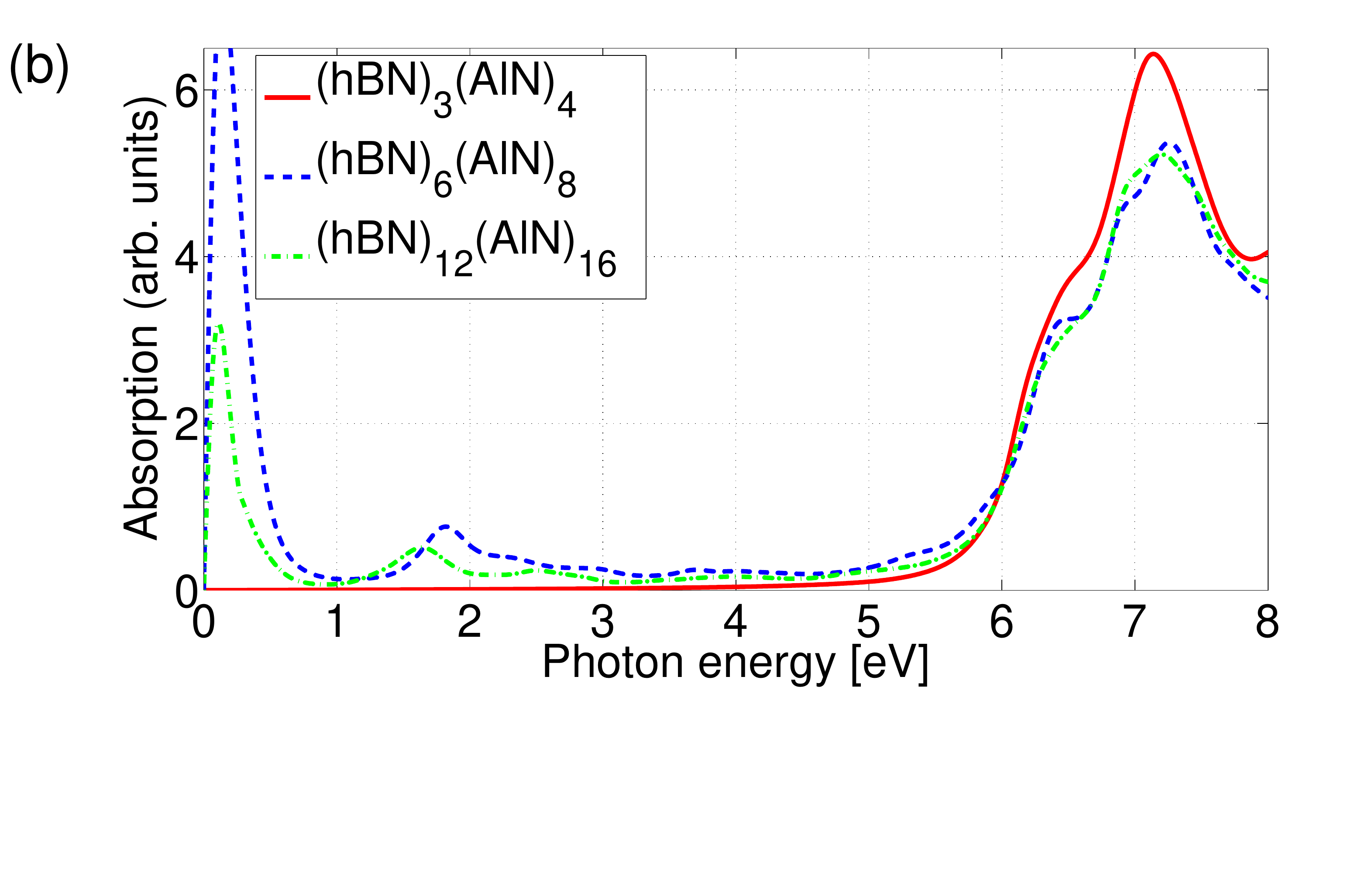}}
\end{center}
\vspace{-0.5cm}
\caption{Optical absorption spectrum of the $(hBN)_3(hAlN)_4$, $(hBN)_6(hAlN)_8$ and $(hBN)_{12}(AlN)_{16}$ SL for light polarization: (a) parallel to [0001] and (b) perpendicular to [0001]. Spectra obtained within the independent particle picture using hybrid-DFT (HSE06) wavefunctions and eigenvalues.}
\vspace{-0.5cm}
\label{absorption}
\end{figure}

We finally turn our attention to the the optical properties of SL.
Fig. \ref{absorption} (a-b) shows the absorption spectrum of the ultrashort $(hBN)_3(AlN)_4$ SL for light polarization parallel and perpendicular to the 
interface. The optical edge is $\approx 5.2$ eV for light polarization along [0001]. For light polarized parallel to the interface the optical gap is $\approx 5.8$ eV, slightly smaller than the values $\gtrsim6$ eV predicted by hybrid-DFT for hBN, hAlN or wAlN bulk materials (see Fig. S7 in Suppl. Mat.). 
Fig. \ref{absorption} (a-b) also shows the calculated absorption spectra for the larger SL $(hBN)_6(AlN)_8$ and $(hBN)_{12}(AlN)_{16}$ SL. 
One notes the formation of a Drude-like peak  for light polarization parallel to the interface, with contribution from transitions between the valence states pinned by the Fermi level at interface A, consistent with the formation of charge carriers free to move along this interface. 
At intermediate energies, optically allowed transitions originating from defect states at interface B result in relatively small-intensity peaks in the optical range at $\sim 2-3$ eV. The weight of these peaks decreases with increasing $L$ increases but remains significant for the layer thickness range considered in this work ($L<10$) nm. At higher photon energies $>5$ eV we remark bulk like features insensitive to $L$, with corresponding optical transitions taking place inside hBN or wAlN layers. The main prominent features in the absorption spectra are situated above $6$ eV in agreement with similar features taking place in bulk materials. These findings suggest that tailorable optical features can be engineered via SL with layers of various thicknesses.

In summary, we have explored the potential of hBN-AlN SL for tailorable properties relevant to UV-LED and photodetector applications. We find via {\it ab initio} calculations that the structure of the AlN layer belongs to the hexagonal phase in ultrashort layers ($\sim 1$ nm) and wurtzite otherwise. This gives rise to a wide range of electronic and optical properties. Depending on the layer length, the electronic properties can vary from insulating behavior with type II band alignment in ultrashort SL to 2D metallic behavior at interfaces in larger SL. SL also display a rich variety of optical properties, with optical gaps as small as $5.2$ eV in ultrashort SL and optically active defect states in the visible range in thicker ones. These unique properties suggest that experimental efforts to realize SL may be worthwhile. 

See Supplementary Material for additional information on the properties of SL, bulk hBN and bulk AlN.

Sandia National Laboratories is a multimission laboratory managed and operated by NTESS, LLC., a wholly owned subsidiary of Honeywell International, Inc., for the U.S. Department of Energy National Nuclear Security Administration under contract DE-NA-0003525.

\newpage
\clearpage
\noindent
\includegraphics[
    page=1,
    width=\textwidth,
    height=\textheight,
    keepaspectratio]{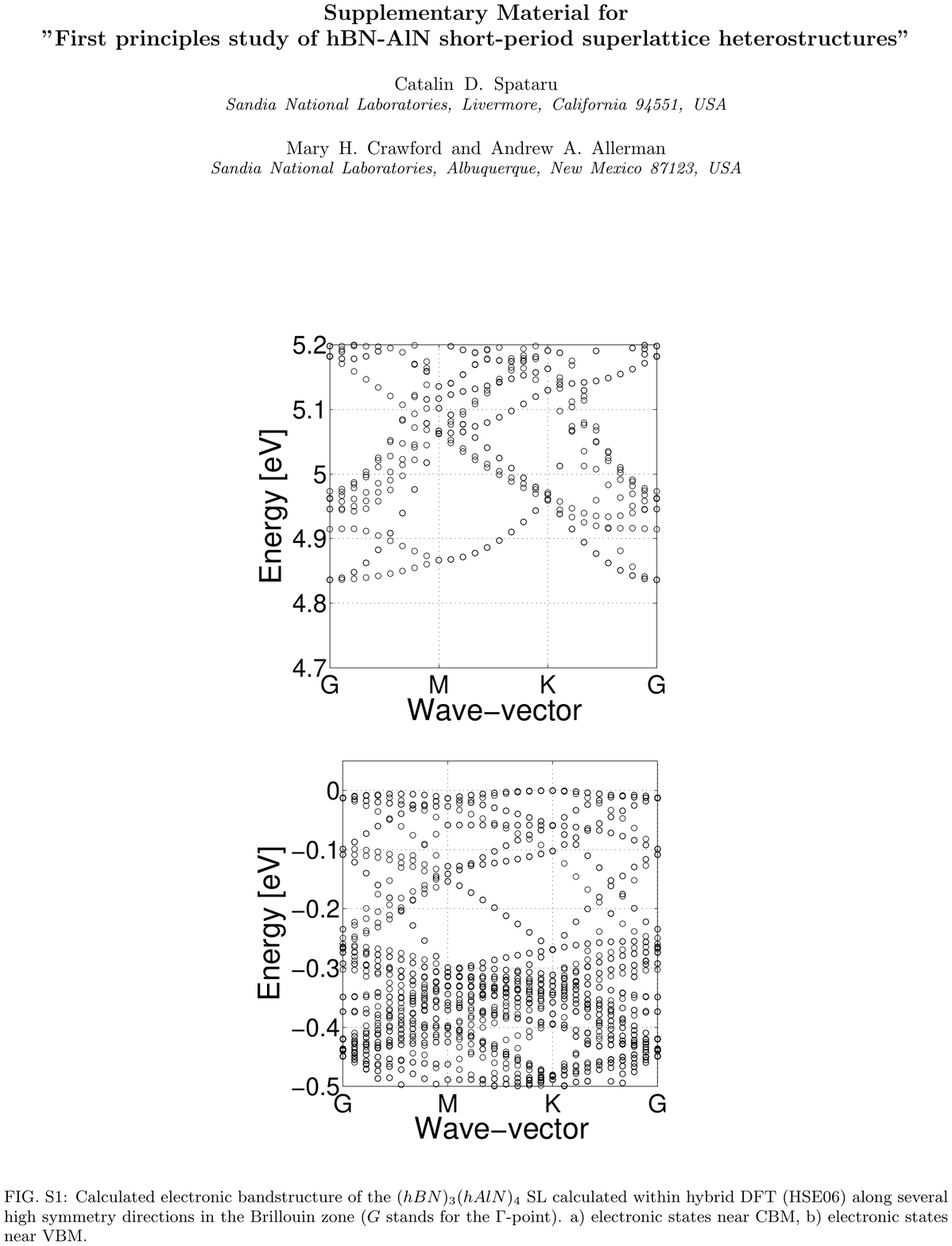}
\vfill
\newpage
\clearpage
\noindent
\includegraphics[
    page=2,
    width=\textwidth,
    height=\textheight,
    keepaspectratio]{Suppl_Mat.pdf}
\vfill
\newpage
\clearpage
\noindent
\includegraphics[
    page=3,
    width=\textwidth,
    height=\textheight,
    keepaspectratio]{Suppl_Mat.pdf}
\vfill
\newpage
\clearpage
\noindent
\includegraphics[
    page=4,
    width=\textwidth,
    height=\textheight,
    keepaspectratio]{Suppl_Mat.pdf}
\vfill
\newpage
\clearpage
\noindent
\includegraphics[
    page=5,
    width=\textwidth,
    height=\textheight,
    keepaspectratio]{Suppl_Mat.pdf}
\vfill
\newpage
\clearpage
\noindent
\includegraphics[
    page=6,
    width=\textwidth,
    height=\textheight,
    keepaspectratio]{Suppl_Mat.pdf}
\vfill

\end{document}